\documentclass{emulateapj}

\usepackage{epsfig}
\usepackage{amsmath}
\usepackage{chngpage}
\usepackage{multirow}
\usepackage{array}
\usepackage{graphicx}
\usepackage{epsfig, graphics}
\usepackage{bbm}
\fontsize{15}{20}
\long\def\symbolfootnote[#1]#2{\begingroup%
\def\thefootnote{\fnsymbol{footnote}}\footnote[#1]{#2}\endgroup} 

\begin{document}

\title{Chandra observations and classification of AGN-candidates \\ correlated with Auger UHECRs}

\author{William A. Terrano}  
\affil{Department of Physics \& CENPA, University of Washington, Seattle, WA 98195, USA}
\author{Ingyin Zaw}
\affil{New York University Abu Dhabi, Abu Dhabi, UAE}
\author{Glennys R. Farrar}
\affil{Center for Cosmology and Particle Physics \&
Department of Physics\\ New York University, NY, NY 10003, USA}

\begin{abstract}
We report on \emph{Chandra} X-ray observations of possible-AGNs which have been correlated with Ultra-high Energy Cosmic Rays (UHECRs) observed by the Pierre Auger Collaboration. Combining our X-ray observations with optical observations, we conclude that one-third of the 21 Veron-Cetty Veron (VCV) galaxies correlating with UHECRs in the first Auger data-release are actually not AGNs. We review existing optical observations of the 20 VCV galaxies correlating with UHECRs in the second Auger data-release and determine that three of them are not AGNs and two are uncertain. Overall, of the 57 published UHECRs with $|b|>10^\circ$, 22 or 23 correlate with true AGNs using the Auger correlation parameters. 
We also measured the X-ray luminosity of ESO139-G12 to complete the determination of the bolometric luminosities of AGNs correlating with UHECRs in the first data-set. Apart from two candidate sources which require further observation, we determined bolometric luminosities for the candidate galaxies of the second dataset. We find that only two of the total of 69 published UHECRs correlate with AGNs (IC5135 and IC4329a) which are powerful enough in their steady-state to accelerate protons to the observed energies of their correlated UHECRs.
The GZK expectation is that $\sim 45$\% of the sources of UHECRs above 60 EeV should be contained within the $z<0.018$ volume defined by the Auger scan analysis, so an observed level of 30-50\% correlation with weak AGNs is compatible with the suggestion that AGNs experience transient high-luminosity states during which they accelerate UHECRs.

\end{abstract}
\maketitle

\section{Introduction}

Identifying the sources of ultra-high energy cosmic rays (UHECRs) is one of the major outstanding goals in astrophysics.  Progress has proven difficult due to the rarity of high energy events and because the deflections of the cosmic rays as they travel through Galactic and extragalactic magnetic fields mean that the cosmic ray arrival directions do not point back to their origins.  The very most energetic cosmic rays ($E \gtrsim 6\times10^{19}$ eV) have attracted attention as a promising path forward.  The energy loss due to the GZK mechanism means that CRs with such energies typically have traveled about 100 Mpc or less, significantly limiting the possible sources for such high energy particles. Furthermore, at these energies magnetic deflections of protons may be small enough to allow the identification of the progenitor type based on statistical associations.

By 2007, the Pierre Auger Collaboration had compiled enough events to begin drawing statistical correlations between UHECR events and possible sources, and reported a strong correlation between UHECRs and the nearby galaxies listed in the \citet{VCV} Catalog of Quasars and Active Galactic Nuclei ($12^{th}$ Ed.) \citep{augerScience07,augerLongAGN}.  The scan procedure that was used steps through UHECR energy threshold, maximum angular separation, and maximum VCV galaxy redshift to find the values giving the lowest chance probability compared to an isotropic distribution, then evaluates the likelihood of such a correlation occurring by chance, by performing the same analysis on many isotropic datasets.  In the following, the term ``correlated" referring to a galaxy with respect to a UHECR simply means that the given galaxy falls within the angular and redshift limits Auger proposed based on applying the scan method to the original dataset using the VCV galaxy catalog.

Of the 27 cosmic rays with energies above 57 $\times 10^{18}$ eV that Auger detected before Aug. 31, 2007, twenty correlate within  $3.2^{\circ}$ with VCV galaxies having $z \leq 0.018$, about 75 Mpc \citep{augerLongAGN}.  There are 21 VCV galaxies correlated with these 20 UHECRs.  (More than one galaxy can be correlated with a UHECR and vice versa.)  Galaxy catalogs such as VCV are incomplete in the Galactic Plane, so a better comparison is obtained by restricting to $|b|>10^{\circ}$ where the VCV catalog is more complete.  With this restriction there are 22 UHECRs of which 19 UHECRs correlate with 20 galaxies \citep{zfg09}.  This is a much higher correlation than would be expected by chance from an isotropic source distribution, and higher than can be explained by nearby galaxy clustering alone \citep{zbf10}. More recent Auger data continue to show a significant, albeit less strong, correlation with VCV galaxies \citep{augerUpdate2010}.  

	
 An important question is whether the observed correlation with VCV galaxies implies that AGNs are the sources of some or all UHECRs. The VCV catalogue is a list of AGN \emph{candidates}, so first it must be established whether the correlating galaxies (i.e., the VCV galaxies within 3.2$^\circ$ of a UHECR) are actually AGNs.  \citet{zfg09} looked at existing observations of the galaxies in VCV that were correlated with the first set of UHECRs and found that only 14 of the 21 galaxies show unambiguous evidence of AGN activity.  Three show no signs of AGN activity, while the other four have ambiguous optical spectra. A widely used technique for optical identification of AGNs, so-called BPT line ratios, compares the relative strengths of diagnostic spectral lines \citep{bpt1981}.  The BPT line ratios of the 4 ambiguous galaxies fall within the ranges classified by \citet{Kauffmann2003} as AGNs but they do not fall within the line-ratio ranges adopted by \citet{KewleyRatio} as indicative of an active nucleus. AGN activity can often be obscured in the optical bands by dust obscuration, and the Kewley test excludes some known AGNs.  Additionally, as many as half of AGNs that are selected based on radio or X-ray properties would not be identified by looking only at their BPT line ratios \citep{reviglioHelfand06}.  Thus, determining whether these 4 ambiguous VCV galaxies have active nuclei requires observations outside of optical wavelengths.

The VCV galaxies are not the only population of nearby galaxies found to correlate with the Auger UHECRs. 
\citet{bzf10} performed an independent scan analysis between the Auger galaxies and nearby Luminous IR galaxies in the PSCz catalog and found an excess correlation.  Restricting to $|b|>10^\circ$ where PSCz is complete, 13 galaxies of $L_{IR} > 10^{10.5} L_{sun}$ correlate within $2.1^{\circ}$ with one or more of the 22 Auger UHECR events.  Some LIRGs contain an active nucleus with dust absorbing the AGN radiation and re-emitting it thermally, giving rise to large IR luminosities. This raises the question of whether the LIRGs found to correlate may actually be AGNs. \citet{bzf10} showed that 6 of the 13 correlating IR galaxies are also in VCV and 5 of these do in fact host AGNs\footnote{The one which is not a confirmed AGN falls within the VCV sample to be tested.}. The other 7 correlating IR galaxies lacked the observations needed to differentiate between star-formation and obscured nuclear activity.

A key distinguishing feature of active galaxies is that accretion-driven radiation produces a broad spectrum extending from the IR to the X-ray, known as the broadband continuum.  Normal and starburst galaxies, on the other hand, have broadened blackbody spectra due largely to stellar emission, which is peaked sharply in the UV/optical. This means the X-ray to optical flux ratios are considerably larger for AGNs than for normal/starburst galaxies.  This has been seen in detailed spectroscopic studies of \emph{Chandra} Deep Field sources \citep{Barger2003}. If the active nucleus is obscured, the X-ray to optical ratio becomes even larger since dust absorbs UV and optical photons more readily than X-rays \citep{Comastri2003}. This makes X-ray observations, particularly when combined with observations in the near-IR, optical or UV, a powerful tool for identifying all classes of AGNs \citep{Maccacaro1988}.

In this paper we use X-ray observations to determine whether the 4 ambiguous VCV galaxies, and the 7 indeterminate PSCz galaxies, have active nuclei.  We observed ten of these possible UHECR source galaxies using the \emph{Chandra} X-ray satellite, while for IC 5169 we used data from a recent XMM-Newton observation.

Another interesting question is whether AGNs that correlate with UHECRs have any characteristic features which distinguish them from the AGNs that do not seem to correlate with UHECRs.  A particularly relevant property is the luminosity of the AGN, as this helps to constrain the possible cosmic ray acceleration mechanisms \citep{fg09}.  Previously, reliable luminosities have been established for all but one of the correlating AGNs for UHECRs in the first data-release\citep{zfg09}, \citep{bzf10}.  We observed the remaining AGN, ESO 139-G12,  in order to obtain the first robust estimate of its bolometric luminosity.

	\begin{table*}
		\begin{center}
		\caption{Table listing source properties}
		\setlength{\tabcolsep}{0.04in}
		{\small
		\begin{tabular}{l p{2cm}  p{1.8cm} p{1.8cm}p{1.8cm}ccp{1.5cm}}
		\hline
		Source Name & Total Counts \newline \scriptsize{ (0.5-10 keV)} & Hard Counts \newline \scriptsize{(2-10 keV)} & Flux \newline (Direct) \scriptsize{ erg cm$^{-2}$ s$^{-1}$} & Flux \scriptsize{(webPIMMS)} \scriptsize {erg cm$^{-2}$ s$^{-1}$} & Rmag & $Log(f_x/f_R) $ &  $L_{2-10}$  \newline \scriptsize {erg s$^{-1}$} \\
		\hline
		IC 5169& 17 & 6 & 3.3 E-14 & 1.4 E -14 & 12.5$^1$ & -3.0 & 7.4 E 39\\
		NGC 7591 & 17 & 2 & 6.6 E -15 & 1.4 E - 14 & 12.4$^{3,b}$ & -3.4 & 8.6 E 39 \\
		NGC 1204 & 13 & 4 & 1.3 E -14 & 1.2 E -14& 13.6$^4$ & -2.9 & 5.8 E 39\\
		NGC 2989 &  5 & 2 &2.3 E -14 & 4.9 E -15 &  12.5$^{1,2}$ & -3.1 & 8.8 E 39\\
		\hline
		\hline
		IC 4523&  6 & 5 & 6.5 E -14 & 6.5 E -15 & 13$^2$  & -2.5  & 3.8 E 40\\
		ESO 270-G007 & 14 & 5 & 2.2 E -14 & -- & 13$^1$ & -3 & 8.4 E 39\\
		IC 5186 & 5 & 0 & 5.8 E -15$^a$ & -- & 12.3$^1$ & -3.8  & 3.4 E 39\\
		ESO 565-G006 & 21 & 8 & 5.5 E -14 & 2.4 E -14 & 12.7$^1$ & -2.7 & 3.2 E 40\\
		NGC 7648 & 5 & 1 & 3.0 E -15 & -- & 12.2$^{3}$ & -4.4 & 9.7 E 38\\
		2MASX J1 754-60 & 3 & 1 & 2.7 E -15 & -- & -- & -- & 1.6 E 39\\
		IC 5179 & 2685$^c$ & -- & -- & 9.1 E -14 & 11.4$^1$ & -2.98 & 2.5 E 40\\
		\hline
		\hline
		ESO 139-G12 & 1070 & 666 & 4.7 E -12$^d$ & -- & 12.9$^{1,2}$ & -0.2 & 2.9 E 43\\
		\hline
		\end{tabular}}
		\label{SourceProp}
	\tablecomments{Fluxes are for 2-10 keV. The first group of four rows are the correlating galaxies found in VCV.  The following six are those from PSCz.  The last row of this group, IC 5179, was observed by XMM-Newton.  Only the direct flux determination method is used for diffuse sources. The final row,  ESO 139-G12, is a known AGN whose  X-ray luminosity had not previously been measured. The 2-10 keV luminosity, $L_{2-10}$, is derived from the direct flux measurement; it is the total X-ray luminosity in the central region and thus an upper limit on X-ray luminosity of a possible AGN, a criterion for the ability to accelerate UHECRs.  \\ $(a)$ No counts above 2 keV, flux given is for 0.5 - 2 keV band. $(b)$ Johnson magnitude rather than Cousins;  this does not affect the result. $(c)$  25ksec XMM-Newton observation;  counts are from 0.5 - 12 keV. $(d)$ Derived from fit, see Figure 1.\\ }
	\tablerefs{(1) \cite{ESOCAT1989}, (2) \cite{HIPASS}, (3) \cite{Vaucouleurs}, (4) \cite{6dFsurvey}}
\end{center}
\end{table*}

\section{Observations and Data Reduction}

Each target galaxy was observed on the Chandra ACIS-I detector for 5 kiloseconds.  The observations took place between January and August of 2009. Data analysis was performed using the {\tt ciao} data analysis software. An X-ray source was observed at all of the targets; the results are given in Table \ref{SourceProp}.

For the \emph{Chandra} observations, we used two different methods to estimate the X-ray energy flux. The direct method sums the energy received per unit time in the band of interest (2-10 keV) in a 2 arcsec window around the source using the {\tt eff2evt} tool which takes into account both the quantum efficiency and the effective area corrections of the satellite. The flux is typically dominated by higher energy and less frequent events, making this technique vulnerable to Poisson fluctuations when the count rates are small.  The second technique considers only the total counts received from the compact nuclear source, ignoring the energies of the photons, and assumes that the spectrum follows a power law with a spectral index of 2. Then webPIMMS can estimate the corresponding flux, including Galactic extinction from webPIMMS and assuming no intrinsic extinction.  For the galaxies which are diffuse sources, we report only the direct measurement of the flux within 2 arcsec of the galactic center, as an upper limit on the possible nuclear emission. 

We also analyzed the XMM-Newton observation of IC 5179.  Using the source imaging PPS processing provided by XMM, we took the count rate on the EPIC PN-CCD camera (0.5-12 keV) and used webPIMMS to estimate the corresponding 2-10 keV flux of the source.

\section{Analysis and Results}

X-ray to optical flux ratios provide a useful way to classify galaxies according to their nuclear activity. \citet{Barger2002} show that for AGNs, the ratio of the 2-10 keV flux to the R-band magnitude is typically $ 1 > log(f_x/f_R)$\symbolfootnote[2]{$log(f_x/f_R) \equiv log_{10}(f_x) + 5.5 + .4*R_{mag}$ where $f_x$ is the 2-10 keV flux}$ > -1$, while obscured AGNs have $log(f_x/f_R) > 1$.  Normal and starburst galaxies have  $log(f_x/f_R) < -2$, making this a robust way of discriminating AGNs, even obscured ones, from normal and starburst galaxies \citep{Donley2005}.  

\begin{figure*}
\includegraphics[width =  \textwidth]{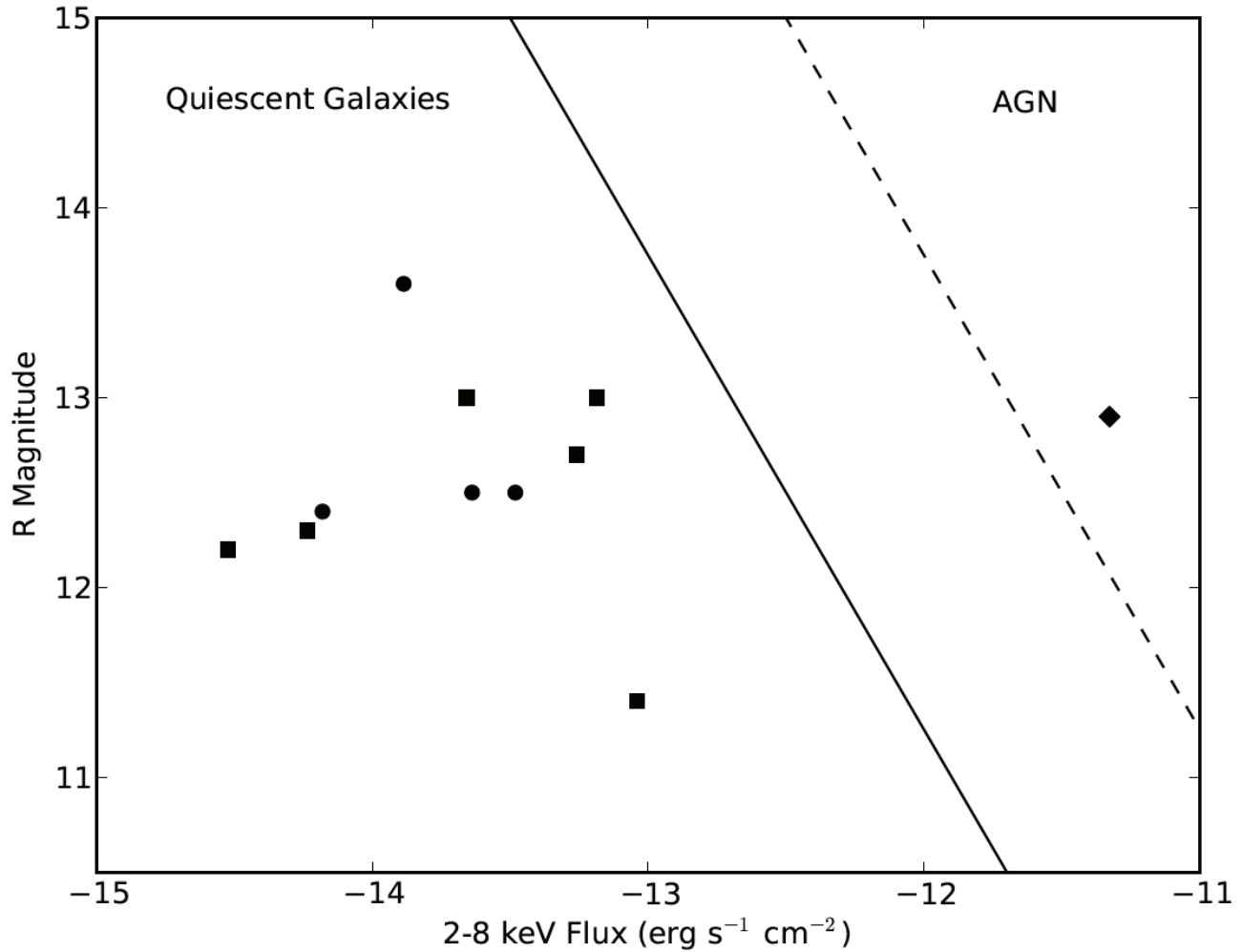}
\caption{\emph{R} magnitude vs. 2-8 keV flux for target galaxies (\textit{squares}: galaxies from the VCV catalogue; \textit{circles}: galaxies from PSCz; \textit{diamond}: ESO 139-G12, the known AGN).  The solid line is $log(f_x/f_R) =  -2$, below which lie quiescent galaxies.  The dashed line is $log(f_x/f_R) = -1$, which separates AGNs on the right from a transition region consisting of Starburst galaxies and weak AGNs.}
\label{OptvsXray}
\end{figure*}

The results are summarized in Figure \ref{OptvsXray}. The X-ray to optical ratios are far below what is expected from AGNs, $log(f_x/f_R) > -1$, for ten of the ambiguous or indeterminate galaxies: the nine we observed with \emph{Chandra} and IC 5179, for which we used an XMM-Newton observation.  This holds even when the larger of the two X-ray flux determinations is used to calculate the X-ray to optical flux ratios. The accuracies of the X-ray flux determinations are limited by shot noise, which at these low counts produces relative uncertainties of order one. Since our measure of nuclear activity depends on $log_{10}(f_x)$, this corresponds to an uncertainty of less than $\sim 0.3$ in $log(f_x/f_R)$. The observed ratios are far from the AGN domain, so our conclusion is robust even with these large uncertainties. We conclude that none of these galaxies have active nuclei. 

The last of the indeterminate galaxies, 2MASX J1 754-60, lacks the optical observations needed for the $log(f_x/f_R)$ test, but its X-ray emission is diffuse and weak, and its near-IR properties are similar to the other galaxies.  We therefore conclude that 2MASX J1 754-60 is unlikely to host an AGN, especially not one expected to be capable of UHECR acceleration.

Obscured AGNs in X-ray surveys are often identified by the hardness ratio of the source. The hardness ratio is defined as H-S/H+S, where H is counts from 2-8 keV and S is counts from 0.5-2 keV. Typically, nearby AGN  with column densities greater than $N_H \approx 10^{22}$ have positive hardness ratios, while unobscured sources have negative hardness ratios \citep{Fiore2000}.  Table \ref{SourceProp} gives counts from  0.5-10 keV and 2-10 keV.  No photons with energies above 8 keV were detected for any of the sources due to the small effective area of \emph{Chandra} at high energies, so these H count values are applicable for estimating the hardness ratio.  Although the hardness ratios have large uncertainties because of the small number of counts, they show no evidence of obscuration.

\section{AGN Identification for the 2nd Auger Data Release}

A second set of UHE events has been released by the Auger collaboration \citep{augerUpdate2010} since the Chandra observations reported here were undertaken.  There are 42 UHECRs in the new sample,  of which 35 have $|b|> 10^\circ$.   Twelve of these (all with $|b|>10^\circ$) correlate within $3.1^\circ$ with 20 VCV galaxies with $z \leq 0.018$.  We have reviewed existing observations of these galaxies to establish which host AGNs.  We could not find published emission lines for two of them, but one (IC 4296) has a radio jet and therefore it has an active nucleus.  The status of the other (G 1314-1532) cannot be determined, but this does not affect the correlation count because another, unambiguous AGN lies within the correlation radius of the same UHECR.  

We use the BPT optical AGN criteria comparing the O[III]/H$\beta$ and N[II]/H$\alpha$ emission line ratios \citep{bpt1981}, to evaluate the nuclear activity in the other 18 correlating VCV galaxies, as was done in \citet{zfg09} for the first Auger UHECRs.  The results are reported in Table \ref{matchedAGN} below.  The classification is clear-cut for 16 of the 18 cases.  For the galaxy Mrk 945, the lower limit of \citet{VO1987} on the O[III]/H$\beta$ line ratio is at the LINER-Seyfert boundary; we note that the actual ratio could easily be significantly larger.   In any case, the Log(S[II] $\lambda 6716 +\lambda 6731$)/(H$\alpha$) ratio of Mrk 945 allows it to be classified as a narrow-line AGN \citep{VO1987}, S[II] being a probe of the partially-ionized region produced by high-energy X-rays from the ionizing source which is a distinguishing feature of active nuclei.  We classify the remaining case, ESO 18-G09, as ambiguous because its emission line ratios satisfy the \citet{Kauffmann2003} but not the \citet{KewleyRatio} criteria.  

Thus we conclude that 3 of the 20 VCV galaxies for the 2nd UHECR dataset do not have active nuclei, while the classification of 2 others cannot be determined with existing data.  
This ``decorrelates" two of the 12 UHECRs but does not change the correlation status of the other three affected UHECRs because those remain correlated with other, established AGN.  
Altogether, as summarized in Table \ref{matchedAGN}, 10 UHECRs with $|b|> 10^\circ$ correlate with confirmed AGNs, one does not, and one remains to be determined.   

We also provide an estimate of the bolometric luminosity, where possible, for correlating AGNs.  From this,  $\lambda_{\rm bol}$ -- a measure of the capability of the AGN to accelerate a proton to the observed energy -- is calculated.   One of the AGNs correlating with a UHECR in the 2nd data-release, \emph{IC4329a}, has a large enough bolometric luminosity to accelerate its UHECR if that is a proton, two are undetermined, and the rest do not.

\begin{table*}
\begin{center}
\caption{New Auger UHECRs and Correlated VCV Galaxies \label{matchedAGN}}
\setlength{\tabcolsep}{0.04in}
{\scriptsize
\begin{tabular}{cccccccccccccc}
\hline
\hline
Year & Day & E$_{CR}$ & RA$_{CR}$ & Dec$_{CR}$ & $l_{CR}$ & $b_{CR}$ & VCV Galaxy & r &  z  & VCVClass &  RealClass & $L_{\rm bol}$ & $\lambda_{\rm bol}$ \\
& & (EeV) & (J2000) & (J2000) & (J2000) & (J2000) & & (deg) & & & & erg s$^{-1}$ & \\
2008 &  87  &  82  &  220.5  &   $-$42.9  &  $-$36.4  & $+$15.5  & NGC 5643  &  2.12 & 0.003 &  S2  &  S2$^1$ & 1.2 $\times$ 10$^{44}$ (a) & 0.18\\
        &      &         &          &           &       &        & IC 4518A  &  2.88 & 0.016 &  S2  &  S2$^2$ & 2.4 $\times$ 10$^{44}$ (b) & 0.36\\
2008 & 192  &  55  &  306.7  &   $-$55.3  &  $-$17.3  & $-$35.4  & IC 4995   &  2.86 & 0.016 &  S2  &  S2$^3$ & 3.1 $\times$ 10$^{42}$ (c) & 0.01\\
2008 & 282  &  61  &  202.3  &   $-$16.1  &  $-$44.0  &   45.8   & MCG -03.34.064 &  1.75 & 0.017 & S1h & S2 or NLS1$^4$ & 2.6 $\times$ 10$^{43}$ (d)& 0.07\\
        &      &         &          &           &       &        & G 1314-1532 & 2.84 & 0.013 &  S?  & ? & --- & --- \\
2008 & 362  &  84  &  209.6  &   $-$31.3  &  $-$40.7  &   29.4   & IC 4329A  &  2.19 & 0.016  &  S1.2 &  S1$^5$ & 1.2 $\times$ 10$^{45}$ (e) & 1.70\\
2009 &  39  &  70  &  147.2  &   $-$18.3  &  $-$106.5 &   26.6   & NGC 2989  &  0.81 & 0.013  &  H2  & H2$^6$ & --- & ---\\
2009 &  51  &  65  &  203.7  &   $-$33.1  &  $-$46.7  &   28.9   & IC 4296   &  0.95 & 0.013  &  S3  & AGN$^7$ & 3.2 $\times$ 10$^{42}$ (f) & 0.01 \\
        &      &         &          &           &       &        & ESO 383-G18 & 0.96 & 0.013 & S1.8 & S2$^4$ & 4.6 $\times$ 10$^{43}$ (g) & 0.11 \\
        &      &         &          &           &       &        & MCG -06.30.015 & 1.22 & 0.008  & S1.5 & S1.2$^8$ & 8.0 $\times$ 10$^{43}$ (h) & 0.19 \\
        &      &         &          &           &       &        & NGC 5253  &  1.82 & 0.001  & H2  & H2$^{9}$ & --- & ---\\
2009 &  78  &  61  &   26.7  &   $-$29.1  &  $-$134.6 &  $-$77.6 & NGC 613   &  2.74 & 0.005  & S?  & Composite$^{10}$ & *** & *** \\
2009 &  80  &  66  &  170.1  &   $-$27.1  &  $-$80.9  &   31.5   & IRAS 11215-2806 & 1.52 & 0.014  & S2 & S2$^{4}$ & 4.0 $\times$ 10$^{43}$ (g) & 0.09\\
2009 &  212 &  57  &  122.6  &   $-$78.5  &  $-$68.8  &  $-$22.8 & ESO 18-G09 &  1.00 & 0.017  & S2 & H2?11 & --- & ---\\
2009 &  219 &  57  &   29.4  &   $-$8.6   &  166.1   &  $-$65.8  & IC 184    &   1.85 & 0.018  & S2 & S2$^{12}$ & 4.8 $\times$ 10$^{43}$ (i) & 0.15\\
        &      &         &          &           &       &        & NGC 788   &   1.98 & 0.013  & S1h & S2 or LINER$^{13}$ & 1.8 $\times$ 10$^{44}$ (j) & 0.55\\
        &      &         &          &           &       &        & IRAS 01475-0740 &  2.21 & 0.017 & S1h & S1h$^{11,14}$ & 2.9 $\times$ 10$^{43}$ & 0.09\\
2009 &  304 &  61  &  177.7  &   $-$5.0   &  $-$83.8  &  54.7    & MCG -01.30.041 &  0.50 & 0.018 & S1.8 & S1.8$^{15}$ & *** & *** \\
2009 &  326 &  57  &    5.4  &   $-$5.6   &   103.3  & $-$67.3   & IRAS 00160-0719 & 1.63 & 0.018 & S2 & H2$^{16}$ & --- & --- \\
        &      &         &          &           &       &        & MRK 945   &   2.43 & 0.015  &  S  & S2$^{17}$ & 5.4 $\times$ 10$^{43}$ (k)& 0.17\\
\hline
\hline
\end{tabular}}
\tablecomments{ {\scriptsize This table lists each Auger UHECR from the update paper (with E $>$ 55 EeV) whose arrival directions are within 3.1$^{\circ}$ of a nearby (z $\leq$ 0.018) VCV galaxy, giving the year and theJulian day each was recorded, their energies and their positions (equatorial and galactic) in degrees \citep{augerUpdate2010}. Next are the correlating VCV galaxies for each, their separation from the UHECR in degrees, and their redshift, and VCV classification. The correct optical classification is then listed (taken from the literature where available);  ``?" indicates the optical classification is ambiguous.  The last two columns show the bolometric luminosity for the AGN and $\lambda_{\rm bol}$, defined as $L_{bol} \times 10^{-45} \times \rm{E_{20}}^{-2}$, the ratio which compares the bolometric luminosity to the theoretical luminosity necessary for producing the associated UHECR. The bolometric luminosities were derived form [OIII] or 2-10 keV luminosities found in literature.\\}}
\tablerefs{{\scriptsize (1) \citet{Wu2011b}, (2) \citet{zfg09}, (3) \citet{Woo2002}, (4) \citet{Gu2006}, (5) \citet{Shu2010}, (6) \citet{NGC2989}, (7) \citet{Cavagnolo2010}, (8) \citet{Bennert2006b}, (9) \citet{Ho1997}, (10) \citet{Veron1997}, (11) \citet{deGrijp1992}, (12) \citet{Kollatschny1987}, (13) \citet{Winter2010}, (14) \citet{Tran2001}, (15) \citet{Kollatschny2008}, (16) \citet{Vader1993}, (17) \citet{VO1987}. (a) \citet{Zhu2011}, (b) \citet{zfg09}, (c) \citet{McKernan2010}, (d) \citet{Wu2011}, (e) \citet{Brightman2011}, (f) \citet{Pellegrini2010}, (g) \citet{Gu2006}, (h) \citet{Vasudevan2009}, (i) \citet{Kollatschny1987}, (j) \citet{Kraemer2011}, (k) \citet{Mulchaey1996}.}}
\end{center}
\end{table*}

\section{Luminosity of ESO 139-G12}

\begin{figure*}
\includegraphics[width = 0.7 \textwidth,angle=270]{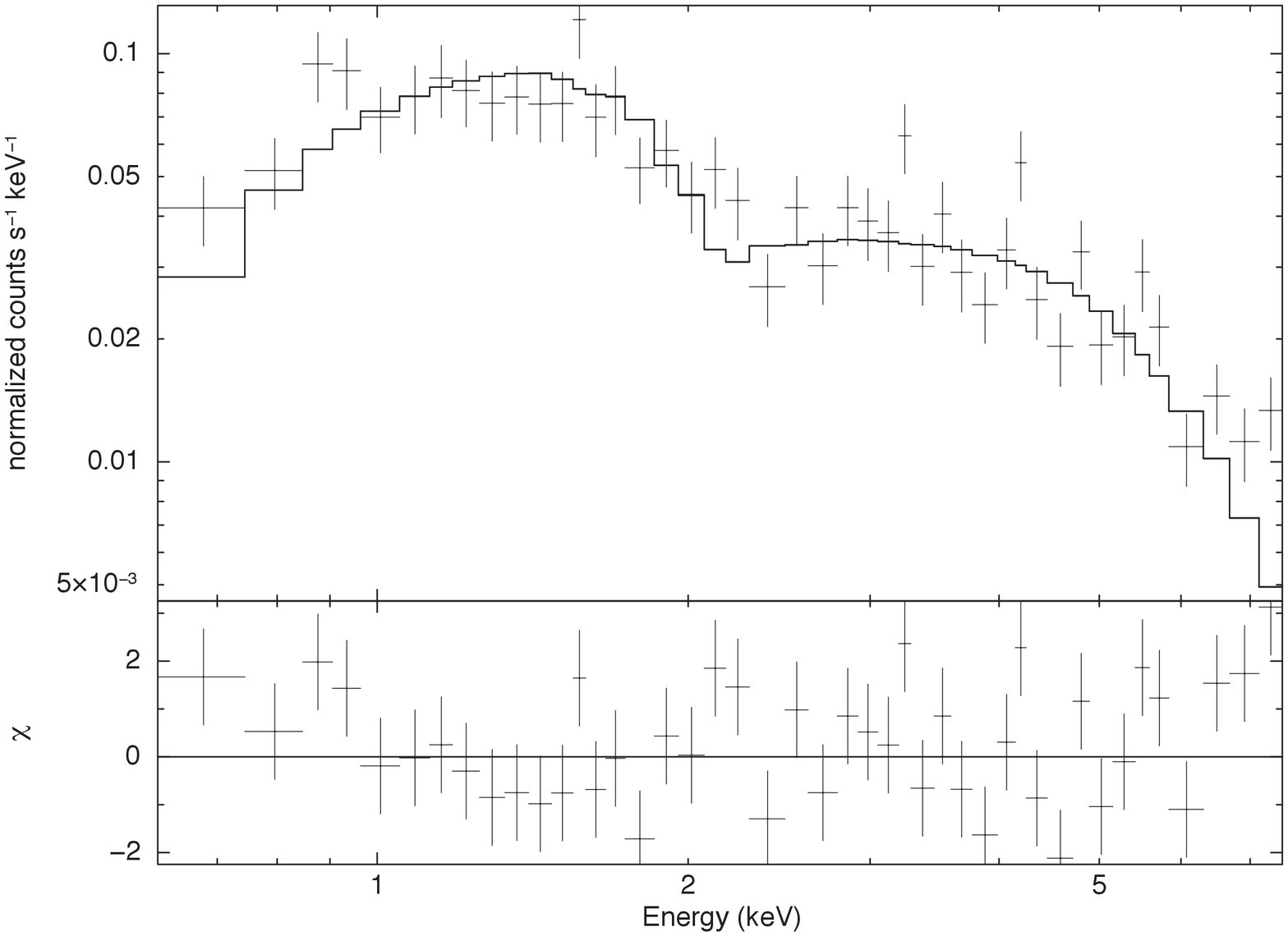}
\caption{Data and fit of the spectrum of ESO 139-G12 to a power law at the source ($\Gamma = 0.72^{+0.09}_{-0.09}$, 90\% confidence) with N$_{\rm H} < 8.2 \times 10^{21} {\rm cm}^{-2}$ at 90\% confidence level.  This is our best fit model of the data, with a reduced $\chi ^2$  of 1.7.  This fit corresponds to a 2-10 keV flux of 4.7$^{+0.6}_{-1.1} \times 10^{-12}$ erg cm$^{-2}$ s$^{-1}$ (error denotes 90\% confidence interval).}
\label{fit}
\end{figure*}

The AGN ESO 139-G12 is correlated with two UHECRs, making it a particularly interesting source candidate.  Previously only an upper limit on its bolometric luminosity existed, derived by \citet{zfg09}.  This source is bright enough for us to determine its flux by fitting the spectrum using {\tt xspec}.  Assuming an absorbed power law, this yields a hard photon index of 0.72$^{+0.09}_{-0.09}$ (error denotes 90\% confidence interval) and very low absorption:  N$_{\rm H} < 8.2 \times 10^{21} {\rm cm}^{-2}$ at 90\% confidence level (Figure \ref{fit}). From this fit we extract a 2-10 keV flux of 4.7$^{+0.6}_{-1.1} \times 10^{-12}$ erg cm$^{-2}$ s$^{-1}$ (error denotes 90\% confidence interval).  When combined with its redshift, $z = 0.017$ ($\approx 74$ Mpc assuming a standard cosmology with $H_o = 70$), we find the 2-10 keV luminosity of ESO 139-G12:  $L_{2-10} \approx 3.1^{+0.4}_{-0.7} \times 10^{42} \, {\rm erg  \  s^{-2}}$.  

The conversion factor relating $L_{2-10}$ to the bolometric luminosity depends on the relative activity of the AGN \citep{LxConv}.  In the absence of a black hole mass measurement for this galaxy, we follow \citet{zfg09} and take $L_{\rm bol} = 20 \, L_{2-10}$, as is typical for an AGN with Eddington ratio less than 0.1.  We therefore estimate the bolometric luminosity of ESO 139-G12 to be $L_{\rm bol} \approx 6.2 ^{+0.8}_{-1.4} \times 10^{43} \, $ erg s$^{-1}$.

For completeness, Table \ref{SourceProp} gives the 2-10 keV luminosity for all of the galaxies which were observed.   With an appropriate conversion factor, this can be used to get an upper limit on possible weak nuclear activity which may be of interest in other studies of UHECR sources.

\section{Discussion}

These \emph{Chandra} observations complete the task of determining which galaxies found to correlate with UHECRs by \citet{augerScience07} and \citet{bzf10} have active nuclei, and of determining the bolometric luminosities of the AGNs correlating with UHECRs in the first data-release.
The X-ray fluxes of the ten unclassified correlating galaxies studied here all fall within the range typical of normal and starburst galaxies of the same optical magnitude, hence we find no evidence of activity in any of them.  Of course, it is impossible to rule out the possibility of very low luminosity nuclear activity which is energetically dominated by the host galaxy, but such weakly active AGNs are not well-motivated candidates for UHECR acceleration anyway \citep{fg09}.

The result, then, is that only 14 out of the original 27 Auger UHECRs (13 out of 22 UHECRs with $|b| \geq 10^{\circ}$) correlate with an actual AGN, using the Auger scan parameters to correlate UHECRs with VCV galaxies but discarding candidate sources which are not in fact AGNs.  Of the 13 highly luminous IR galaxies ($L_{\rm IR} \geq 10^{11.5} \, L_\odot$) found by \citet{bzf10} to correlate within $2.1^\circ$ with one (or more) of the 22 UHECRs with $|b| \geq 10^{\circ}$, we find that only five are also AGNs.\footnote{We could not run this correlation for the full 27 UHECRs because IRAS does not observe within the Galactic plane.}  Thus five of the 27 UHECRs, and one of the 22 UHECRs with $|b| \geq 10^{\circ}$, have neither an AGN nor a LIRG within 3.2$^\circ$ or $2.1^\circ$ respectively.   

From the second Auger data set \citep{augerUpdate2010}, we find 10 UHECRs with $|b|> 10^\circ$ correlate with confirmed AGNs, one does not, and one remains to be determined. 

Thus using the full dataset of UHECRs with $|b|>10^\circ$, where Galactic extinction does not hide source candidates, $(13+10) / (22 + 35) =  0.40^{0.51}_{0.32}$ of the UHECRs correlate with a confirmed AGN using the correlation parameters proposed in the original Auger analysis \citep{augerScience07}, where the upper and lower values are $1-\sigma$ limits for Poisson statistics \citep{gehrels}.  It is interesting that when actual AGNs rather than VCV galaxies are used, the agreement between the correlation found in the early and later Auger datasets becomes closer.  Using the correct galaxy attributions found here, the first and second data sets individually give correlations to confirmed AGNs of $13/22 = 0.59^{0.80}_{0.43}$ and $10/35 = 0.29^{0.41}_{0.20}$ which differ at the 1.07-$\sigma$ level, compared to the individual datasets being 1.63-$\sigma$ from the mean using the uncorrected VCV-attribution.  If the last VCV galaxy is confirmed as an AGN, the correlation would rise to $(13+11) / (22 + 35) =   0.42^{0.53}_{0.34} $ of the UHECRs correlating with an AGN, and $11/35 = 0.31 ^{0.44}_{0.22}$ in the second dataset alone.

Given the uncertainty in the fraction of VCV galaxies which are actually AGNs, and the fact that the scan parameters were determined prior to removing non-AGNs, it is difficult to assess the significance of the final correlation.  Further complicating correlation studies is the incompleteness of source catalogs (especially within the Galactic plane), the composition uncertainty, and magnetic deflection that can obliterate the angular correlation between the CRs and their true source,  particularly for UHECRs with charge $Z >1$.  Indeed, the new Galactic magnetic field model of \citet{jf12}, with a more general form for the field constrained by extensive, all sky RM and polarized synchrotron emission data, predicts that Galactic deflections are small in some portions of the sky but are large in others, even for protons.

The one established AGN which we observed in order to determine its bolometric luminosity, ESO 139-G12, proves to have $L_{\rm bol}$ comparable to most of the other 13 correlated AGNs examined in \citet{zfg09}.  Knowledge of $L_{\rm bol}$ and the energies $E \equiv E_{20} \,10^{20}$eV of the correlated UHECRS (89 and 59 EeV) allows us to evaluate $\lambda_{\rm bol} \equiv 10^{-45} L_{\rm bol} \, E_{20}^{-2}$, the figure-of-merit introduced by \citet{zfg09} to quantify the ability of an AGN to accelerate a proton to the energy of the correlated UHECR.  A value of $\lambda_{\rm bol} \gtrsim 1$ satisfies the acceleration criterion for protons (c.f., \citet{fg09}).  With $\lambda_{\rm bol} \sim 0.1$, ESO 139-G12 is thus marginal according to standard UHECR acceleration mechanisms for protons.

GZK energy losses imply that (taking sources to be uniformly distributed in redshift) about $45$\% of the sources of protons above 60 EeV should have $z<0.018$.  Thus -- taking at face value the 30-50\% correlation we find between UHECRs and confirmed (albeit weak) AGNs -- all UHECRs could have been produced in galaxies presently hosting (generally weak) AGNs, consistent with the picture of transient production of UHECRs via exceptional, powerful flares flares in very weakly or non-active galaxies \citep{fg09}.  Examples of tidally produced flares in quiescent galaxies have recently been discovered in archival SDSS data \citep{vf11} and observed (apparently in blazar mode) by the \emph{Swift} satellite \citep{Burrows2011}, \citep{Bloom2011}.  The spectral energy distribution of the \citet{vf11} flares are well-fit by a thin accretion disk model and the resultant bolometric luminosities amply satisfy the minimum luminosity requirement for UHECR acceleration in both cases (GRF, in preparation);  depending on how rapidly the evidence of the accretion episode disappears, the host galaxy of a tidal disruption flare may or may not show evidence of weak AGN activity in later observations such that it would appear in a catalog such as VCV.  It is also possible, given the uncertainties, that other candidate sources not associated preferentially with AGNs may be responsible for some, most, or all UHECRs.  Indeed, other studies have found correlations between the arrival directions of the Auger UHECRs and a variety of extragalactic sources:  HI galaxies (Ghisellini et al. 2008), AGNs from the Fermi catalog (Nemmen et al. 2010), Swift-BAT AGNs and 2MRS galaxies \citep{augerUpdate2010}.  

\section{Summary}
Our observations and analysis using the \emph{Chandra} X-ray satellite and other data establish that one-third of the 21 galaxies in the Veron-Cetty Veron catalog of AGN candidates found to correlate with UHECR arrival directions in the first Auger data release \citep{augerScience07}, do not in fact have active nuclei.  Combining this with our measurement of the X-ray luminosity of ESO 139-G12, an AGN correlating with two UHECRs, implies that only one of the 27 UHECRs in that first Auger data-release correlates with an AGN (IC 5135) which is powerful enough in its steady-state to accelerate protons to the observed energies, according to conventional acceleration mechanisms \citep{fg09}.  None of the correlated PSCz source candidates we observed have active nuclei.

We reviewed optical observations of the correlating VCV galaxies for UHECRs in the second data release to identify which galaxies actually have an active nucleus -- 3 do and two are ambiguous -- and to determine their bolometric luminosities.  For one of them, IC 4329a, $L_{\rm bol}$ is high enough to accelerate a proton to the observed energy in the steady state; $L_{\rm bol}$ is not determined for two AGNs, and is insufficient in the remaining cases.

Combining the first and second Auger data releases and adopting the original Auger scan parameters, our results are compatible with 30-50\% of UHECRs being produced by genuine but weak AGNs within $z<0.018$.  A consistent picture is that a significant portion of UHECRs are produced in the flaring state of otherwise very weak AGNs \citep{fg09}, since about 45\% of the sources of UHE protons should have $z<0.018$.  However until much larger UHECR datasets and more complete and pure catalogs of potential source candidates become available, the source(s) of the highest energy cosmic rays will not be settled. 

This research was supported by NASA through Chandra Award No. GO9-0130X.   In addition, the research of GRF has been supported in part by NSF-PHY-0701451 and NSF-PHY-0900631.   GRF and IZ acknowledge their membership in the Pierre Auger Collaboration and thank their colleagues for helpful discussions.  We are grateful to S. Veilleux and R. Yan for help clarifying the proper identification of MRK 945.




\clearpage

\end{document}